\documentclass[journal]{IEEEtran}
\usepackage{amsmath,amsfonts}
\usepackage{algorithmic}
\usepackage{algorithm}
\usepackage{array}
\usepackage[caption=false,font=normalsize,labelfont=sf,textfont=sf]{subfig}
\usepackage{textcomp}
\usepackage{stfloats}
\usepackage{url}
\usepackage{verbatim}
\usepackage{graphicx}
\usepackage{cite}
\usepackage{booktabs}
\usepackage{tabularx}
\usepackage{makecell}
\begin{document}

\title{QoE-Aware Service Function Chain Orchestration in Space–Air–Ground Integrated Networks for In-Flight Internet}

\author{Feng Liu, Bing Du, ~\IEEEmembership{Member,~IEEE}, Ziyang Guo, Xiaoqi Zhang, ~\IEEEmembership{Member,~IEEE}, Haijun Zhang, ~\IEEEmembership{Fellow,~IEEE} 
        % <-this % stops a space
\thanks{This study was supported by the National Natural Science Foundation of China under Grant 62471031. (Corresponding author: Bing Du)}% <-this % stops a space
\thanks{Feng Liu, Bing Du, Ziyang Guo, Xiaoqi Zhang, and Haijun Zhang are with the School of Computer and Communication Engineering, University of Science and Technology Beijing, Beijing 100083, China. (e-mail: m202310616@xs.ustb.edu.cn; dubing@ustb.edu.cn; M202420965@xs.ustb.edu.cn; zhangxiaoqi@ustb.edu.cn; haijunzhang@ieee.org).}
}

\markboth{}%
{Liu \MakeLowercase{\textit{et al.}}: QoE-Driven Service Function Chain Scheduling in SAGIN via Multi-Agent Reinforcement Learning}

%\IEEEpubid{0000--0000/00\$00.00~\copyright~2025 IEEE}
%\IEEEpubidadjcol

\maketitle

\begin{abstract}In remote regions lacking terrestrial infrastructure, aviation users increasingly demand Internet services comparable to ground cellular networks. This paper proposes a space–air–ground integrated network (SAGIN) architecture that combines ground networks, civil aviation networks, and low Earth orbit (LEO) satellites. Based on real North Atlantic flight data, a heterogeneous aviation Internet is constructed to support diverse aircraft service requirements.

A composite Quality of Experience (QoE) metric, characterized by transmission delay, jitter, and content similarity, guides efficient resource utilization. The dynamic connectivity of SAGIN is modeled using a reconfigurable time-expanded graph (RTEG). With Network Function Virtualization (NFV), data delivery is formulated as a Service Function Chain (SFC) deployment across ground stations, relay aircraft, and satellites, resulting in a long-term QoE maximization problem.

A tailored Multi-Agent Deep Deterministic Policy Gradient (MADDPG) algorithm incorporating LCSS-based path filtering and softmax-based action selection is developed. Simulation results demonstrate that the proposed approach enables seamless connectivity for heterogeneous aircraft and consistently outperforms game-theoretic and MATD3 methods in terms of QoE.

\end{abstract}


\section{INTRODUCTION}
W{\scshape ith} the rapid evolution of communication technologies, mobile users increasingly demand ubiquitous and seamless connectivity, and air travelers now expect in-flight Internet services comparable to terrestrial cellular networks. Current civil aviation communications mainly rely on ground-based Internet Gateway Stations (IGSs) to establish air-to-ground (A2G) links. However, over oceans and remote regions where A2G coverage is limited or unavailable, air-to-space (A2S) communication becomes the primary means of connectivity. Despite its global coverage, A2S communication is constrained by limited satellite bandwidth and high latency. In addition, the high mobility of aircraft—up to 250 m/s—leads to large-scale, highly dynamic network topologies. Under these constraints, providing reliable, low-latency connectivity comparable to terrestrial networks during transoceanic flights remains a major challenge.

To overcome these limitations, Aerial Ad Hoc Networks (AANETs) have emerged as a promising solution \cite{ref1}. AANETs enable aircraft-to-aircraft (A2A) communication through long-range, highly directional wireless links, allowing aircraft within communication range to form dynamic, flexible, and cost-effective mesh networks. By exploiting aircraft mobility and spatial distribution, AANETs effectively improve access efficiency across the airspace. Accordingly, extensive efforts have been devoted to enhancing AANET communication performance \cite{ref4,ref7,ref8,ref9}. When combined with conventional air-to-ground (A2G) links, A2A communications allow aircraft to relay data to ground gateways via multi-hop paths, thereby extending the effective coverage of ground base stations.

Nevertheless, the limited communication range of relay aircraft may leave some aircraft isolated in sparsely populated airspace, posing challenges to achieving global connectivity solely through AANETs. Satellite networks are therefore incorporated into the AANET framework to mitigate this limitation. Through the synergistic integration of abundant ground network resources, the flexibility and mobility of airborne networks, and the ubiquitous coverage provided by satellite systems, the Space–Air–Ground Integrated Network (SAGIN) architecture is formed. This architecture enables seamless wide-area connectivity for aircraft operating in highly dynamic and heterogeneous environments \cite{ref11, ref12, ref_zhang_SAGIN, ref_zhang_joint, ref_zhang_LEO}.

Building on this paradigm, recent studies have investigated the use of unmanned aerial vehicles (UAVs) as an additional layer for collaborative resource optimization \cite{ref_zhang_joint, ref_UAV1, ref_UAV2}. However, despite their agility, UAVs suffer from inherent constraints such as limited endurance, payload capacity, and onboard energy, which restrict their applicability in civil aviation scenarios. Consequently, increasing attention has been directed toward exploiting the high density and mobility of commercial aircraft themselves to enable mutual assistance. For instance, Hoffmann et al. proposed a genetic algorithm–based framework for joint optimization of gateway allocation, routing, and scheduling in integrated aircraft–ground–satellite networks, aimed to minimize average packet delay \cite{ref13}. Similarly, Guo et al. designed a three-layer architecture that combined satellite and ground-assisted aircraft caching to reduce transmission latency for onboard passengers \cite{ref14}.

Most existing studies evaluate aviation Internet performance using traditional Quality of Service (QoS) metrics, such as latency, throughput, and bit error rate. However, QoS-centric approaches exhibit inherent limitations. They primarily emphasize technical indicators while overlooking user perception and content characteristics, and often rely on rigid resource allocation strategies that may result in low utilization efficiency and poor adaptability to highly dynamic network conditions \cite{ref15}. With the diversification of in-flight services, aviation networks are required not only to support bandwidth-intensive applications such as video streaming but also to accommodate delay-sensitive services like video conferencing, alongside more delay-tolerant applications including web browsing, email, social media, and data synchronization \cite{ref16}. Given the heterogeneous service requirements and varying user tolerance levels, optimization based solely on QoS is insufficient to capture overall system performance. Therefore, this paper advocates the adoption of a more flexible and user-centric metric—Quality of Experience (QoE)—to evaluate and optimize aviation Internet services within the three-layer SAGIN architecture \cite{ref17,ref18,ref19}.

Additionally, the high mobility and heterogeneity of SAGIN introduce significant real-time variability and management complexity. To tackle these challenges, Network Function Virtualization (NFV) is adopted to decouple network functions from underlying hardware, thereby improving resource utilization and management efficiency \cite{ref_SFC}. Within the NFV framework, data delivery from source nodes to destination aircraft can be abstracted as the deployment of a Service Function Chain (SFC), in which each node along a transmission path implements a Virtual Network Function (VNF). Accordingly, routing in SAGIN can be formulated as a multi-task SFC scheduling problem, which is non-convex and NP-hard, motivating the use of reinforcement learning–based solutions.

Existing work, such as \cite{ref20}, combines game theory with actor–critic learning to optimize bandwidth allocation for individual data streams. In contrast, this study focuses on the concurrent servicing of multiple aircraft and aims to derive cooperative SFC decisions that maximize the overall network Quality of Experience (QoE). To this end, a Multi-Agent Reinforcement Learning (MARL) framework is introduced, enabling distributed agents to learn coordinated strategies. Specifically, we employ the Multi-Agent Deep Deterministic Policy Gradient (MADDPG) algorithm, which adopts an actor–critic architecture and exhibits strong convergence and robustness in multi-agent environments \cite{ref30}. An extension of MADDPG is developed to handle discrete and time-varying action spaces arising from dynamic path availability, by incorporating the Longest Common Subsequence (LCSS)-based path filtering together with a softmax mechanism. Each agent outputs a probability distribution over feasible paths, enabling deterministic execution while preserving differentiability during training.

To accurately characterize the highly dynamic topology of SAGIN, a Reconfigurable Time-Extended Graph (RTEG) model is further developed. The aerial layer is constructed from real North Atlantic flight data involving 571 aircraft during peak periods, the satellite layer consists of 66 Iridium LEO satellites, and the ground layer includes 20 coastal gateway stations. These layers are integrated into a unified SAGIN framework, within which aircraft may connect directly to ground stations, via relay aircraft, or through satellites when terrestrial links are unavailable. On this basis, the proposed space–air–ground integrated Multi-Agent Deep Deterministic Policy Gradient (SAG-MADDPG) algorithm adaptively optimizes SFCs deployment for service-demanding aircraft to enhance overall QoE. Finally, the impacts of service types, buffer capacities at ground stations and relay aircraft, and the number of destination aircraft are systematically evaluated, with comparisons against scenarios excluding satellite assistance. For clarity, the main contributions of this work are as follows:
\begin{itemize}
\item We propose a three-tier SAGIN architecture in which the aerial network serves as a functional service layer rather than a supplementary component. By exploiting the dynamic density of airborne networks, the architecture enables direct and opportunistic relay transmissions between aircraft, effectively extending ground coverage. An RTEG model captures the highly dynamic topology, while Network Function Virtualization reformulates resource management as a coordinated SFC deployment problem.
\item QoE is introduced as a performance metric to support diverse service types in the aviation network, allowing flexible and efficient allocation of network resources to better meet passenger needs. The definition of QoE incorporates both objective indicators—such as latency and jitter—and a subjective factor, content similarity.
\item The proposed SAG-MADDPG algorithm, incorporating LCSS-based path filtering and a softmax mechanism, is employed to derive optimal SFCs for aircraft with data demands. Simulation results based on real North Atlantic flight data show that, when the SAGIN scales to approximately 650 nodes, the proposed approach outperforms the Multi-Agent Delay-Deterministic Policy Gradient (MATD3) and game-theoretic methods, achieving faster convergence and superior QoE.
\end{itemize}

The remainder of this paper is organized as follows. Section II presents the network architecture, the reconfigurable time-extended graph model, and the four SFC transmission modes. Section III introduces the communication model, the QoE formulation, and the optimization objective. Section IV outlines the algorithmic framework and describes the reconstruction of the algorithm based on the problem model. Section V evaluates the performance of the proposed method, and Section VI concludes the paper with a summary of the main findings. The main variables and their corresponding definitions used in this paper are summarized in Table \ref{tab:table1}.

\begin{table}[t]

\centering
\caption{Summary of Key Notions\label{tab:table1}}

    \begin{tabular}{c|c}
        \toprule
        Symbol & Definition\\
        \hline
        ${\mathbb{I}}$ & \makecell[l]{Set of $N$ Internet gateway stations (IGS), i.e., \\ $\{ {I_1}, \cdots ,{I_N}\} $}\\
        \hline
        $\mathbb{X}$ & \makecell[l]{Set of $K$ satellites, i.e., $\{ {X_1}, \cdots ,{X_K}\} $}\\
        \hline
        ${\mathbb{V}}$ & \makecell[l]{Set of $M$ air vehicles, i.e., $\{ {V_1}, \cdots ,{V_M}\} $}\\
        \hline
        $M_D$ & \makecell[l]{Number of destination aircraft}\\
        \hline
        $M_R$ & \makecell[l]{Number of relay aircraft}\\
        \hline
        ${\mathbb{V}^D}$ & \makecell[l]{Set of $M_D$ destination aircraft, i.e.,  $\{ {V_1^D}, \cdots ,{V_{M_D}^D}\} $, \\$\mathbb{V}^D \in {\mathbb{V}}$}\\
        \hline
        ${\mathbb{V}^R}$ & \makecell[l]{Set of $M_R$ relay aircraft, i.e., $\{ {V_1^R}, \cdots ,{V_{M_R}^R}\} $,\\  $\mathbb{V}^R \in {\mathbb{V}}$}\\
        \hline
        $\mathbf{G}(t)$ & \makecell[l]{The SGAIN dynamic graph}\\
        \hline
        $\mathbf{RG}(t_n)$ & \makecell[l]{The reconstructed snapshot at $t_n$}\\
        \hline
        $\eta$ & \makecell[l]{The rate of the change of the edges between the \\adjacent subgraphs}\\
        \hline
        ${\tau}$ & \makecell[l]{Delay of the entire transmission process}\\
        \hline
        ${J}$ & \makecell[l]{Jitter of transmission}\\
        \hline
        $\xi$ & \makecell[l]{Similarity of content before and after transmission}\\
        \hline
        $\mathbf{\chi}$ & \makecell[l]{The pre-transmission content eigenmatrix}\\
        \hline
        $\tilde{\mathbf{\chi}}$ & \makecell[l]{The post-transmission content eigenmatrix}\\
        \hline
        $\mathbf{P}_i$ & \makecell[l]{The feasible path selection space for aircraft in $\mathbb{V}^D$}\\
        \hline
        $\mathbf{P}^{\text{dim}}_i$ & \makecell[l]{The number of optional paths for $V^D_i$}\\
        \hline
        $\varrho ^{k-hop}_i$ & \makecell[l]{Total number of $k$-hop path options for $V^D_i$,\\ with $k \in \{1, 2, 3\}$}\\
        \hline
        $\varrho^{S}_i$ & \makecell[l]{The number of optional A2S paths for $V^D_i$}\\
        \hline
        $\varrho^{act}_i$ & \makecell[l]{The serial number of path selected by $V^D_i$}\\
        \bottomrule
    \end{tabular}
\end{table}

\section{SYSTEM MODEL}

In this section, we first present the architecture of the SAGIN. Subsequently, a reconfigurable time-expanded graph-based model of the SAGIN is constructed using real trans-North Atlantic flight trajectory data. Finally, we introduce the four SFC transmission modes supported within the SAGIN framework.

\subsection{Network Infrastructure}

\begin{figure}[ht]
    \centering
    \includegraphics[width=1.0\linewidth]{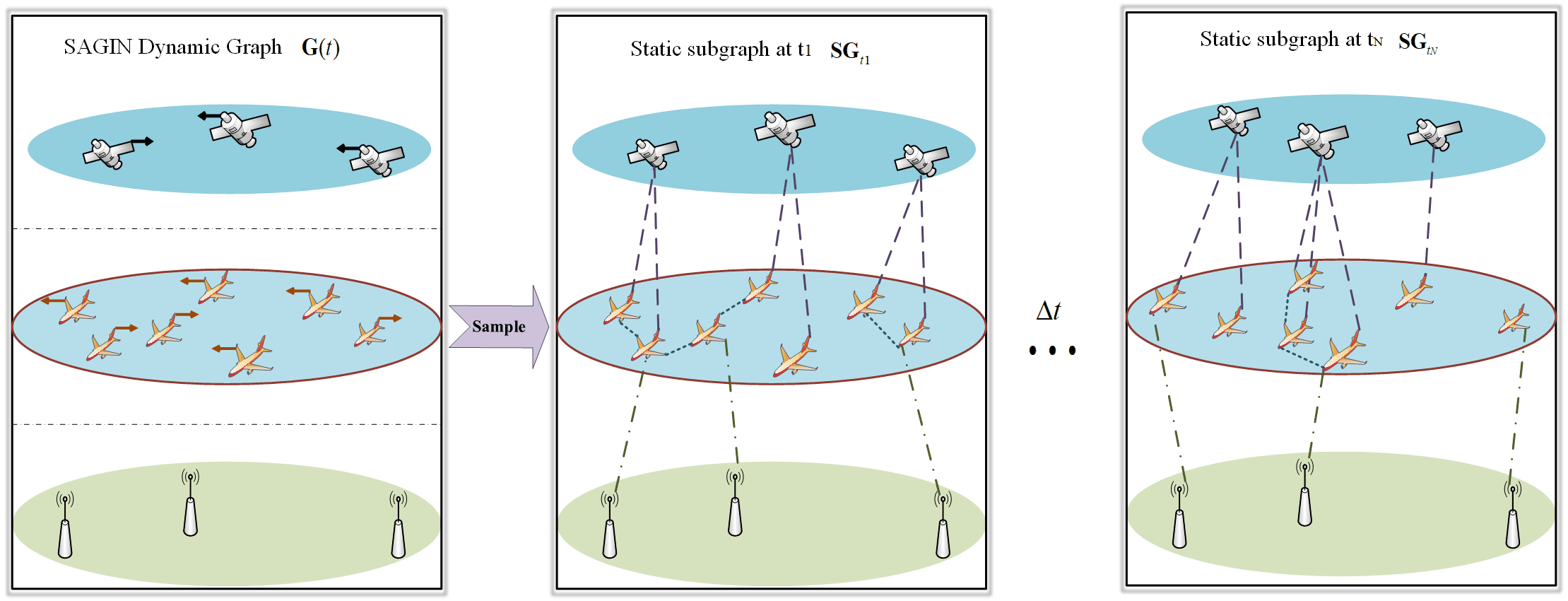}
    \caption{~~SAGIN Reconfigurable Time-Expanded Graph}
    \label{fig:dynamic-graph}
\end{figure}

The architecture of the SAGIN is illustrated in Fig. \ref{fig:dynamic-graph}. The satellite layer comprises LEO satellites from the Iridium constellation, denoted by the set \( \mathbb{X} = \{X_1, \cdots, X_k, \cdots, X_K\} \), where \( K \) is the total number of satellites. The aerial layer consists of civil aviation aircraft cruising at high altitudes, represented by the set \( \mathbb{V} = \{V_1, \cdots, V_m, \cdots, V_M\} \), with \( M \) denoting the number of aircraft. The ground layer includes Internet gateway base stations deployed across terrestrial locations, defined as \( \mathbb{I} = \{I_1, \cdots, I_n, \cdots, I_N\} \), where \( N \) is the total number of base stations.

Within the SAGIN framework, aircraft can establish three types of communication links:
\begin{itemize}
\item{}Air-to-Ground (A2G) enables communications between aircraft and ground base stations. While ground stations offer abundant communication resources, their coverage radius is limited to approximately 200 nmi due to hardware constraints \cite{ref21}.
\item{ }Air-to-Air (A2A) supports direct communications between aircraft, with a typical transmission range of 250 nmi. Aircraft in the network can not only exchange information but also act as relays for others. This capability extends the effective coverage of A2G links and enables flexible, dynamic networking based on the mobility of aircraft. In addition, A2A communication generally incurs lower costs compared to satellite-based links.
\item{}Air-to-Space (A2S) offers near-global coverage and ensures consistent connectivity for all aircraft. LEO satellites in the Iridium constellation orbit at an altitude of approximately 780 km, and the latency is in the range of 20-250 ms \cite{ref32}.
\end{itemize}
By fully leveraging the complementary strengths of these three communication modes, the proposed SAGIN architecture aims to deliver reliable, flexible, and cost-effective connectivity for aircraft, thereby enhancing in-flight service quality for passengers while ensuring robust communication across the airspace.

\subsection{Reconfigurable Time-Expanded Graph}

Given the highly dynamic nature of SAGIN, we adopt a reconfigurable time-expanded graph (RTEG) to model its time-varying topology. RTEG represents the evolving network as a compact sequence of graph snapshots by identifying critical topology-change events \cite{ref_RTEG}. In SAGIN, such events mainly arise from link establishment or disruption, typically caused by nodes entering or leaving communication ranges or by link quality crossing operational thresholds.

Aircraft positions are predicted in real time using actual flight trajectory data and combined with known locations of ground gateway stations and LEO satellites. This enables continuous inference of the dynamic network topology, denoted by $\mathbf{G}(t) = \{\mathbf{V}(t), \mathbf{E}(t)\}$, where $\mathbf{V}(t)$ includes satellites, aircraft, and ground stations, and $\mathbf{E}(t)$ denotes the set of available wireless links.

A set of critical time instants ${t_1,t_2,\ldots,t_N}$ is defined, each corresponding to one or more topology-change events. The network is assumed quasi-static between adjacent critical instants. Accordingly, a sequence of reconstructed snapshots $\{\mathbf{RG}_{t_1},\mathbf{RG}_{t_2},\ldots,\mathbf{RG}_{t_N}\}$ is generated, where each snapshot $\mathbf{RG}_{t_n}$ represents a static topology over the interval $\{t_n,t_{n+1}\}$. These snapshots are then temporally connected to form the complete RTEG model.

RTEG construction includes:

\begin{itemize}
    \item Node replication, where each physical node $v$ is replicated as $(v,t_n)$ at every critical time point;
    \item Spatial edges, which reflect valid connectivity within each snapshot and are weighted by propagation distance;
    \item Temporal edges connect the replica nodes $(v,t_n)$ and $(v,t_{n+1})$ of the same physical node, capturing the continuity of node states across adjacent time intervals and the cost of waiting or local persistence.
\end{itemize}

To efficiently identify critical time points, we introduce a topology change sensitivity metric $\eta$, defined as the ratio of cumulative link additions and deletions since the last critical time $t_n$ to the number of edges in $\mathbf{RG}_{t_n}$:

\begin{equation}
\label{eq1}
\eta(t) = \frac{N_E(\mathbf{RG}(t)-\mathbf{RG}(t_n))}{N_E(\mathbf{RG}(t_n))}\tag{1},
\end{equation}

 A threshold $\eta_{th}$ (e.g., 1\%) is applied: when $\eta(t) \geq \eta_{th}$, a new critical time point is created and a new snapshot is reconstructed. An overly small threshold leads to redundant snapshots, while an excessively large one may miss significant topology changes. Proper tuning of $\eta_{th}$ therefore balances modeling accuracy and computational complexity.

\subsection{SFC Transmission Modes}
The SAGIN architecture provides a unified hardware and connectivity platform for Network Function Virtualization (NFV). In this framework, ground Internet gateway stations $\mathbb{I}$ and satellites $\mathbb{X}$ act as data sources, while aircraft with data demands serve as destination nodes $\mathbb{V}^D \subset \mathbb{V}$. Aircraft without data requests operate as relay nodes $\mathbb{V}^R \subset \mathbb{V}$, forwarding data toward destinations. Virtual Network Functions (VNFs) are deployed on relay aircraft, ground gateways, and satellites, forming Service Function Chains (SFCs), where each SFC is an ordered set of VNFs that together provide an end-to-end network service.

\begin{figure}[ht]
    \centering
    \includegraphics[width=1.0\linewidth]{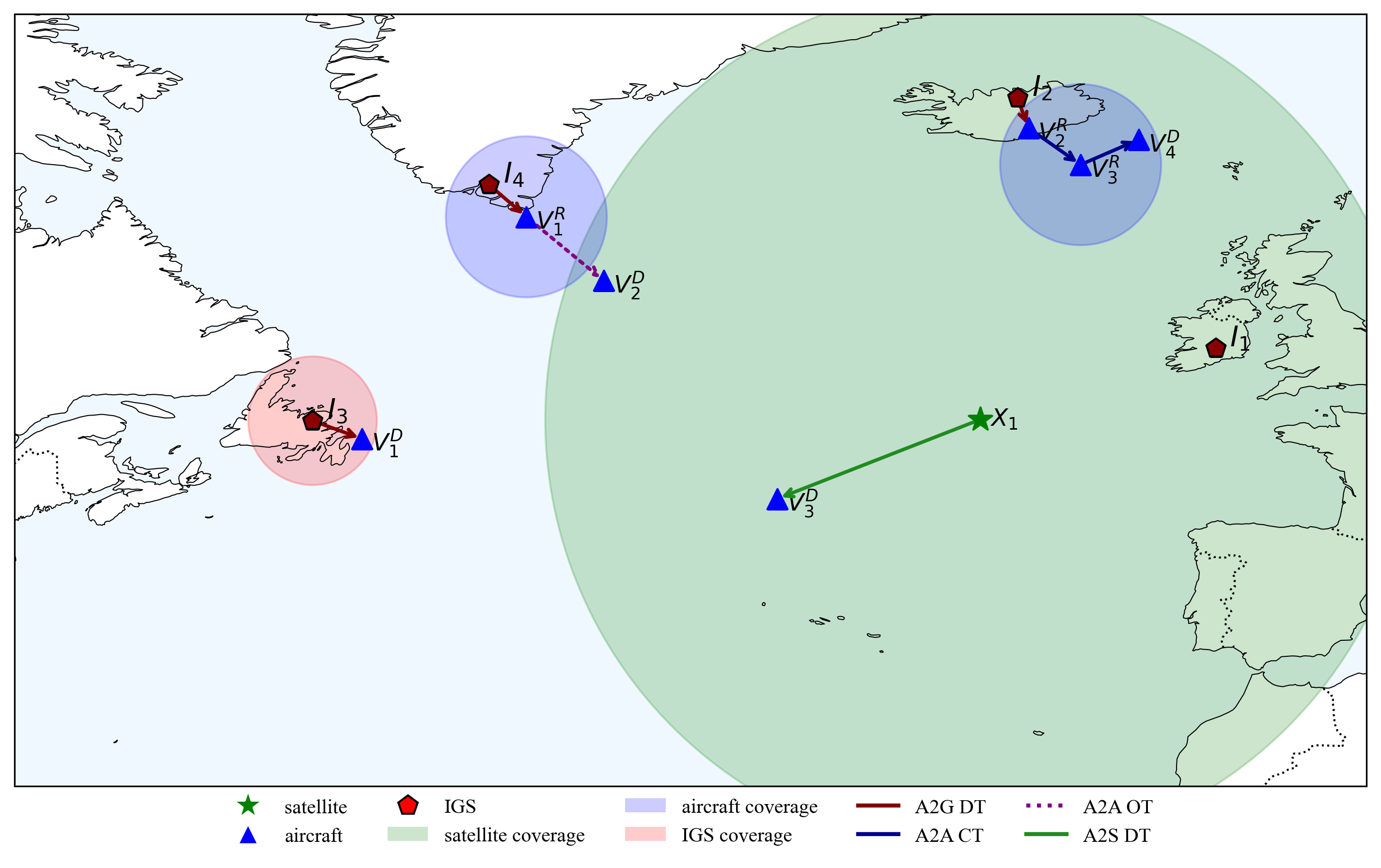}
    \caption{~~Overview of the Four SFC Transmission Modes in SAGIN}
    \label{fig:coverage-transmode}
\end{figure}

Figure \ref{fig:coverage-transmode} illustrates the feasible SFC transmission modes within the SAGIN network. As shown, aircraft $V_1^D$ is located within the coverage area of ground station $I_3$, enabling air-to-ground direct transmission (A2G-DT). In this case, SFC1, $I_3 \to V^D_1$, is completed within a single subgraph. Aircraft $V_4^D$ can communicate with $I_2$ via relay aircraft $V_3^R$ and $V_2^R$, supporting air-to-air connection transmission (A2A-CT). Accordingly, SFC2, $I_2 \to V^R_2 \to V^R_3 \to V^D_4$, is also completed within one subgraph.

For aircraft $V_2^D$, when neither A2G nor A2A links are available, data can be transmitted from ground station $I_4$ to relay aircraft $V_2^R$ for temporary caching until the SFC $V_2^R\to V_2^D$ becomes available. This transmission mode corresponds to air-to-air opportunistic transmission (A2A-OT). Consequently, SFC3 $I_4 \to V^R_1 \to V^D_2$ spans adjacent subgraphs and is constrained by buffer capacity and delay tolerance.

In scenarios where both ground stations and relay aircraft are unavailable, as illustrated by $V_3^D$, data are delivered via satellite links, enabling air-to-space direct transmission (A2S-DT). In this case, SFC4 is given by $X_1 \to V^D_3$.

\section{NETWORK MODEL AND PROBLEM FORMULATION}

In this section, we first introduce the communication network model based on the SAGIN architecture. Then, the proposed Quality of Experience (QoE) model is presented. Finally, a QoE-based utility function is formulated to capture the objective of SAGIN SFC optimization at the network level.

\subsection{Communication Model}

In the SAGIN network, both A2A and A2G communications are assumed to operate in the VHF band and share spectral resources, whereas A2S communications are conducted in the L-band \cite{ref11}. The A2A, A2S and A2G communications are essentially at high altitude, so both can be approximated as Line-of-sight (LOS) propagation links \cite{ref23}. The network consists of $N$ IGSs, $K$ satellites, and $M$ aircraft, including $M_D$ destination aircraft and $M_R$ relay aircraft, where $M_D+M_R<M$. This paper focuses primarily on downlink transmissions from ground stations or satellites to destination aircraft. The following sections use the A2G downlink as a representative example for illustration.

The distance of A2G link connecting the vehicle  ${V^D_i}$ and the base station ${I_n}$ is denoted as $d_{i,n}^{{\rm{A2G}}}$.  The channel gain of an A2G $G_{i}^{{\rm{A2G}}}$ LOS link can be calculated by the free-space path loss model, and the path loss exponent is taken as 2. 

\begin{equation}
G_{_{i}}^{{\rm{A2G}}} = \frac{{{g_T^I}{g_R^V}{c^2}}}{{(4\pi f{\kern 1pt} d_{_{i,n}}^{{\rm{A2G}}}){\,^2}}} \tag{2}, \label{eq2}
\end{equation}
where ${g_T^I}$ denotes the transmitting antenna gain of the IGS, ${g_R^V}$ represent the receiving antenna gain of the air vehicle, ${c}$ is the speed of light, ${f}$ is the carrier frequency.

The $p_t^{I_n}$ represents the transmit power of IGS ${I_n}$. According to \cite{ref29}, the signal-to-interference-plus-noise ratio (SINR) is calculated as follows:

\begin{equation}
{\text{SINR}_i} = \frac{{{p_t^{I_n}}{G_i}}}{{ {\textstyle \sum_{k\in{\mathcal{C}^{int}_i }}}p_t^k{G_k}+ {\sigma ^2}}} \tag{3},
\label{eq3}
\end{equation}
where ${\sigma ^2}$ is the noise variance. Let $\mathcal{C}^{int}_i$ be the set of transmitters within the interference range of aircraft $V^D_i$, defined by the maximum distance threshold $d^{int}_{max}$. The total interference is then the sum of the transmission powers from all nodes in this set. Then the maximum achievable transmission rate ${r_i}$ can be derived by:

\begin{equation}
\label{eq4}
r_i = W{\log _2}(1 + {\text{SINR}_i}) \tag{4}.
\end{equation}
In (\ref{eq4}), ${W}$ represents the total bandwidth between the transmitter and receiver.

During the flight, the destination aircraft $V^D_i$ establishes a SFC with the ground base station via A2A or A2G, with the help of VNFs, to obtain the required data resources. Let ${\textbf{SFC}}_{i} = \left\{ {{I_n},V^R_1, \cdots ,{V_\vartheta^R },{V_i^D}} \right\}$ define the path ${I_n}\to {V^R_1}\to \cdots \to{V_\vartheta^R}\to{V^D_i}$, $\vartheta$ is the number of relay VNFs in this path. The delay at node $V^D_i$ equals the total delay across all hops along the path:

\begin{equation*}
\label{eq5}
\tau_i= {\textstyle \sum_{v\in{\textbf{SFC}'_i}}^{}} \frac{L_{v{-},v}}{B_v^{ach}} + \delta T \tag{5},
\end{equation*}
where ${L_{v^{-},v}}$ represents the length of the data transmitted between two nodes. $v^{-}$ denotes the predecessor of node $v$ in ${\textbf{SFC}}_{i}$, i.e., the node responsible for forwarding the required data to aircraft $v$. $\textbf{SFC}'_i$ refers to the subset of nodes in ${\textbf{SFC}}_{i}$ excluding the IGS, ${\textbf{SFC}}'_{i} = \left\{ {V^R_1, \cdots ,{V_\vartheta^R },{V_i^D}} \right\}$. Due to the limited coverage and data resources of the base stations and the relay nodes, the aircraft may not have a feasible path at all times, so ${\delta}$ is the number of time slots for which $V^D_i$ has been waiting for transmission. ${T}$ represents the length of a time slot.

Depending on ${\tau_i}$, the jitter experienced by the receiver can be defined as follows:

\begin{equation}
\label{eq6}
{J_i} = \left| {{\tau _i}(t) - {\tau _i}(t - 1)} \right|\tag{6}.
\end{equation}
It refers to the variation in the delay of the user in two adjacent transmissions.

\subsection{QoE Model}

The relationship between latency and QoE exhibits a negative correlation characterized by a distinct threshold effect. When latency is low, QoE remains high and stabilizes; however, once latency exceeds a certain threshold, QoE experiences a significant decline. Similarly, the relationship between jitter and QoE is also negatively correlated, demonstrating an approximately exponential decay. 

Additionally, content similarity is commonly employed to assess the quality of user experience. The definition of similarity varies based on the form of the transmitted content: for text transmission, it refers to semantic similarity \cite{ref24}, while for images and videos, it is typically measured using the Structural Similarity Index Measure (SSIM) \cite{ref25}. The fundamental principle behind these metrics involves extracting features from the pre-transmission and post-reception content, and then calculating the similarity between these features. Therefore, in this paper, content similarity $\xi_i $ is defined as follows:

\begin{equation}
\label{eq7}
\xi_i =\frac{\mathbf{\chi} _{i}(\tilde{\mathbf{\chi} }_{i} ) ^{T }  }{|| \mathbf{\chi} _{i} ||\cdot  || \tilde{\mathbf{\chi} }_{i} ||  } \tag{7}, 
\end{equation}
where $\mathbf{\chi} _{i} $ denotes the pre-transmission content eigenmatrix, $\tilde{\mathbf{\chi} }_{i}$  represents the post-transmission content eigenmatrix,  $( \bullet ) ^{T}$ indicates the transpose of a matrix, and \( \left \| \bullet  \right \|  \) denotes the $l2$ norm of a matrix. $\xi_i \in \left [ 0,1 \right ] $, the similarity is maximized when $\xi_i = 1$, indicating no loss or degradation of resources during transmission, whereas $\xi_i = 0$ corresponds to the opposite scenario. It can be inferred that $\xi_i$ exhibits a nonlinear positive correlation with QoE: when $\xi_i$ exceeds a certain threshold $\xi^{th}$, QoE increases sharply, while below this threshold, QoE declines gradually and steadily, stabilizing at $q_{low}$, and $q_{low}\in(0,1)$\cite{ref25,ref26}.

The QoE of the aircraft $V^D_i$ can be formulated by

\begin{align*}
{\text{QoE}_i} & = 5\cdot  \big(\frac{\rho_1}{{(1 + {e^{_{{\theta _1}({\tau _{i}} - {\tau ^{th}})}}})}} + \frac{\rho_2}{{e^{{\theta _2}{J_i}}}} \\& + \rho_3\left [q_{low}+ \frac{(1-q_{low})}{e^{{\theta _3}({\xi ^{th} -\xi_i }) }} \right ] \big) \tag{8}. \label{eq8}
\end{align*}
In (\ref{eq8}), ${\tau ^{th}}$ denotes the latency threshold, ${\theta _1}$ and ${\theta _2}$ respectively represent the positive attenuation coefficients associated with delay and jitter, and ${\theta _3}$ denotes the growth rate of similarity. The parameters \( \rho_1 \), \( \rho_2 \), and \( \rho_3 \) represent the weighting coefficients corresponding to the contributions of latency \( \tau \), jitter \( J \), and content similarity \( \xi \) to the receiver’s QoE, respectively. These coefficients satisfy the constraint \( \rho_1 + \rho_2 + \rho_3 = 1 \). According to (\ref{eq8}), the gain of the whole network can be expressed as

\begin{equation}
\label{eq9}
{\cal U} = \sum\limits_{i = 1}^{M_D} {{\text{QoE}_{i}}} \tag{9}.
\end{equation}

\subsection{Problem Formulation}
The objective of this study is to enable each aircraft with data requests (i.e., the destination nodes in the RTEG $\mathbf{G}(t)$) to select an SFC that meets predefined constraints, such that the overall QoE across the SAGIN network is maximized. The objective function can be written as

\begin{equation}
\label{eq10}
\max \ \mathcal{U} = \max \ \sum\limits_{i = 1}^{M_D} {{\text{QoE}_i}} \tag{10}.
\end{equation}
\begin{subequations}
\begin{align}
{\mathrm { s.t.}} \ & {\textstyle \sum_{k \in \mathcal{C}^{ca}_n}} L_{n,k} \le \text{BS}^{max}_n\tag{10a}\label{eq10a}\\ & \tau_i\le \tau^{max} \tag{10b}\label{eq10b}\\ & 0 \le \text{QoE}_i\le 5 \tag{10c}\label{eq10c}   
\end{align} 
\end{subequations}
In (\ref{eq10a}), $\mathcal{C}^{ca}_n$ denotes the set of aircraft that need to store data at node $n$. Equation (\ref{eq10a}) enforces the buffer-capacity constraint for relay aircraft and IGSs in SAGIN, requiring the cached data of each VNF to remain within its maximum buffer limit $\text{BS}^{\max}$. Equation (\ref{eq10b}) ensures that the transmission delay of the selected chain for each destination node remains within its maximum tolerable delay $\tau^{max}$. Equation (\ref{eq10c}) defines the permissible range of the QoE for each destination node.

\section{ALGORITHM}

This section begins with an overview of the basic MADDPG algorithm, followed by a Markov Decision Process (MDP) based formulation of the problem, and concludes with the introduction of the proposed SAG-MADDPG algorithm.

\subsection{MADDPG Algorithm}
The Multi-Agent Deep Deterministic Policy Gradient (MADDPG) algorithm extends DDPG from single-agent continuous control to multi-agent environments. In multi-agent settings, non-stationarity caused by simultaneous policy updates degrades learning performance. MADDPG addresses this issue by adopting a centralized training and decentralized execution (CTDE) paradigm \cite{ref27}. During training, each critic accesses global information to evaluate joint actions, whereas during execution, each actor selects actions based solely on local observations, ensuring scalability and practicality.

In this section, ``$M$" denotes the number of agents, ``$m$" indexes a specific agent, ``$\theta$" represents network parameters, and ``$\tau$" is the soft-update rate; these notations differ from those used previously.

\subsubsection{Actor-Critic Architecture}
Each agent $m$ maintains an actor network $\pi_m(o_m(t),|,\theta^{\pi_m})$ and a critic network $Q_m(\mathbf{s}(t),\mathbf{a}(t),|,\theta^{Q_m})$, along with corresponding target networks $\pi'_m$ and $Q'_m$. The actor maps the local observation $o_m(t)$ to a deterministic action $a_m(t)$, while the critic evaluates the Q-value using the global state $\mathbf{s}(t)={o_1(t),\ldots,o_M(t)}$ and the joint action $\mathbf{a}(t)={a_1(t),\ldots,a_M(t)}$.

Target networks are updated via soft updates to stabilize training, and experience replay is employed to improve sample efficiency. While critics exploit global information during training, execution remains fully decentralized.

\subsubsection{Training Process of Actor Network}

Training samples are stored in a centralized replay buffer as tuples $\langle \mathbf{s}(t),\mathbf{a}(t),\mathbf{r}(t),\mathbf{s}(t+1)\rangle$, where $\mathbf{r}(t)={r_1(t),\ldots,r_M(t)}$. Each agent optimizes its policy by maximizing the expected Q-value estimated by its critic:

\begin{equation}
    \nabla_{\theta^{\pi_m}} \mathcal{J}_m = \mathbb{E}_{\mathbf{s}} \left[ \nabla_{\theta^{\pi_m}} \pi_m(o_m(t)) \cdot \nabla_{\hat{a}_m} Q_m(\mathbf{s}(t), \hat{\mathbf{a}}(t)) \right]. \tag{11}\label{eq11}
\end{equation}

For a mini-batch of $N_{mb}$ samples, gradients are computed per sample and averaged:

\begin{equation}
    \nabla_{\theta^{\pi_m}} \mathcal{J}_m = \frac{1}{N_{mb}} \sum_{k=1}^{N_{mb}} \nabla_{\theta^{\pi_m}} \mathcal{J}_m^k. \tag{12}\label{eq12}
\end{equation}
Finally, the actor parameters are updated by gradient ascent:

\begin{equation}
    \theta^{\pi_m} \leftarrow \theta^{\pi_m} + \beta \cdot \nabla_{\theta^{\pi_m}} \mathcal{J}_m. \tag{13}\label{eq13}
\end{equation}

\subsubsection{Training Process of Critic Network}

For each sampled transition, target actions are generated by target actor networks, and the target Q-value is computed as:

\begin{equation}
     y_m(t)=r_m^k(t)+\gamma Q'(\mathbf{s}_m^k(t+1),\hat{\mathbf{a'}}_m^k(t+1)|\theta ^{Q'_m}). \tag{14}\label{eq14}
\end{equation}
The temporal-difference (TD) error is the difference between the current critic's estimate and the target value:

\begin{equation}
     \delta _m^k = Q(\mathbf{s}_m^k(t),\hat{\mathbf{a}}_m^k(t)|\theta ^{Q_m}) - y_m^k(t). \tag{15}\label{eq15}
\end{equation}
The critic is trained by minimizing the mean squared TD error:
\begin{equation}
    \theta ^{Q_m} \leftarrow \theta ^{Q_m}-\alpha \cdot \nabla_{\theta^{Q_m}} \mathcal{L}_m. \tag{16} \label{eq16}
\end{equation}
where $\mathcal{L}*m=\frac{1}{2N*{mb}}\sum_k(\delta_m^k)^2$.

To stabilize learning, all target networks are updated through a soft update mechanism, where their parameters are periodically synchronized with the main networks using a small update rate \( \tau \),

\begin{equation}
 \theta_{\text{target}} \leftarrow \tau \theta + (1-\tau)\theta_{\text{target}}. \tag{17} \label{eq17}
\end{equation}

\subsection{Problem Reformulation}

\begin{figure*}[htb]
  \centering
  \includegraphics[width=0.75\linewidth]{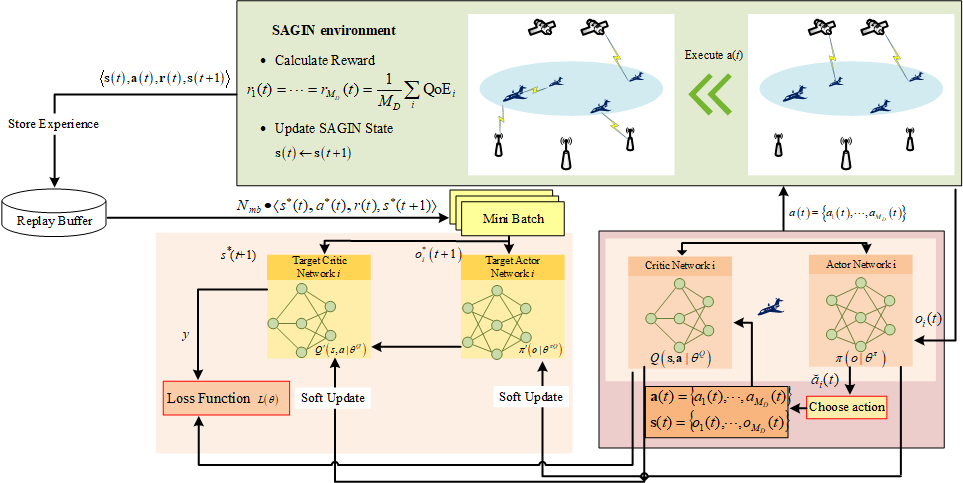}
  \caption{~~SAG-MADDPG Algorithm}
  \label{fig:maddpg}
\end{figure*}

Within the proposed framework, every destination aircraft $V^D_i \in \mathbb{V}^D$ generating a data request is represented as an autonomous agent that determines its SFC by leveraging observable network states and past learning experience. Each destination aircraft is assumed to issue a request once every three time slots. Given the predictable aircraft positions, the location of each aircraft at each time slot can be estimated. Accordingly, the feasible chain space $\mathbf{P}_i$ for each $V^D_i$ is constructed from subgraphs $\{ \text{RG}_{t_1}, \text{RG}_{t_2}, \text{RG}_{t_3} \}$, representing the network topology over three consecutive time slots. Each aircraft selects an SFC from $\mathbf{P}$, and the joint decisions of all aircraft in $\mathbb{V}^D$ determine the final aeronautical internet topology, upon which QoE is evaluated.

To solve the optimization problem using MADDPG, the system is formulated as an MDP, defined by its state space, action space, and reward function, which together describe agent-environment interactions during training.

\subsubsection{State Space}

In the system, there are \( M_D \) destination aircraft. At the beginning of time step $t$, each destination aircraft $V^D_i$, where $i \in (1, M_D)$, acquires its path information from the network, which forms its local observation $o_i(t)$. Each aircraft can only obtain information relevant to itself. Therefore, the global state $s(t)$ at time step $t$ is constructed by aggregating the local observations $o_i(t)$ of all destination aircraft. The definitions of the local observation $o_i(t)$ and the global state $s(t)$ are given as follows.

\begin{equation}
    o_i(t) = \{ \tau_i(t-1), \mathbf{P}_i^{\text{dim}}(t), \varrho_i^{1-hop}, \varrho_i^{2-hop},\varrho_i^{3-hop}, \varrho_i^{S}\} \tag{18}\label{eq18}
\end{equation}
\( o_i(t) \) denotes the local observation of the \( i \)-th destination aircraft at time step \( t \), where
\begin{enumerate}
\item{$\tau_i(t-1)$}: The delay of the agent $i$ at the previous data transmission, .
\item{$\mathbf{P}_i^{\text{dim}}(t)$}: The number of optional paths for the airplane $i$ at time step t.
\item{$\varrho_i^{1-hop}$}: The number of one-hop paths in the optional SFCs, i.e., A2G-DT links.
\item{$\varrho_i^{2-hop}$}: The number of two-hop paths in the optional SFCs containing A2A-CT and A2A-OT links.
\item{$\varrho_i^{3-hop}$}: The number of three-hop paths in the optional SFCs containing the A2A-CT and A2A-OT links.
\item{$\varrho_i^{S}$}: Number of satellite communication SFCs in the optional paths, including A2S-DT links.
\end{enumerate}
Each agent collects its own observation, and the state of the current time step $t$ is available:

\begin{equation}
    \mathbf{s}(t) = \{ o_1(t), \cdots,o_i(t), \cdots, o_{M_D}(t)\} \tag{19}\label{eq19}
\end{equation}
and thus, the state space can be expressed as $S = \{ \mathbf{s}(t),t\in(0,T^{tot})\}$.

\subsubsection{Action Space}

Based on $o_i(t)$, the corresponding action $a_i(t)$ can be obtained:

\begin{equation}
    a_i(t) = \{ \varrho^{act}_i\} \tag{20} \label{20}
\end{equation}
where $\varrho^{act}_i$ indicates the serial number of the SFC in $\mathbf{P}_i(t)$ selected by destination aircraft $i$. The set of all destination aircraft actions at the current time step $t$ is represented as follows:

\begin{equation}
    \mathbf{a}(t) = \{ a_1(t), \cdots,a_i(t), \cdots, a_{M_D}(t)\} \tag{21}\label{eq21}
\end{equation}
 The action space should be expressed as $A = \{ \mathbf{a}(t),t\in(0,T^{tot})\}$. The output action $a_i(t)$ range is determined by the SFC space size $\mathbf{P}_i^{\text{dim}}(t)$ for each destination aircraft.

\subsubsection{Reward Function}
Once the chains for all destination aircraft are determined, the resulting network topology initiates data transmission, and the QoE for each aircraft is subsequently measured. The primary objective of this study is to maximize the overall QoE of the aviation interconnection network. Since the SFC selection of each aircraft influences not only its own QoE but also that of others, cooperative decision-making among aircraft is essential to optimize the global reward. To enable coordinated learning, all aircraft share a common reward signal, encouraging convergence toward a globally optimal policy. The reward function for aircraft $i$ is defined as follows:

\begin{equation}
    r_i(t) = \frac{1}{M_D} \sum\limits_{m = 1}^{M_D} {{\text{QoE}_m}}\tag{22} \label{eq22}
\end{equation}
The reward for each aircraft is the average of the QoE of all destination nodes. And the reward for the current time step $t$ is then $\mathbf{r}(t) = \{r_1(t), \cdots, r_i(t), \cdots, r_{M_D}(t)\}$.

\subsubsection{Discrete Action Space Adaptation}
In this study, the action space is discrete and varies across aircraft and time slots, making discrete MARL methods with fixed action dimensions (e.g., MADQN) unsuitable \cite{refMADQN}. To overcome this limitation, we adopt an LCSS-based SFC filtering mechanism and integrate it with a softmax-based action selection strategy, thereby enabling the continuous-action MARL framework MADDPG to operate effectively in the proposed discrete and variable action space. 

Specifically, the actor network outputs a real-valued vector, which is transformed into a probability distribution over feasible SFCs using the softmax function. This probabilistic representation enables smooth gradient propagation during training while maintaining compatibility with discrete decision-making. To accommodate the variable action space, we define a maximum action dimension $\mathbf{P}_{\max}^{\mathrm{dim}}=F$. The Longest Common Subsequence (LCSS) algorithm is employed to measure structural similarity among candidate chains and filter out highly redundant ones, thereby ensuring sufficient diversity and maintaining high-quality representations within the action space. From the original candidate set $\tilde{\mathbf{P}}_i(t)$, up to $F$ representative chains are retained to form the effective action space $\mathbf{P}_i(t)$, whose dimension is given by $\mathbf{P}_i^{\mathrm{dim}}=\min (\tilde{\mathbf{P}}_i^{\mathrm{dim}},F)$. Accordingly, the actor network produces $\mathbf{P}_i^{\mathrm{dim}}$ logits, which are normalized by the softmax function to obtain a valid probability distribution. If the number of available SFCs is smaller than $F$, only the first $\tilde{\mathbf{P}}_i^{\mathrm{dim}}$ outputs are retained and normalized.

During execution, a deterministic policy is adopted: each aircraft selects the SFC with the highest softmax probability as its final action. In contrast, during training, the softmax formulation preserves differentiability with respect to the actor parameters, enabling efficient policy-gradient updates within the MADDPG framework. In this way, the proposed design bridges continuous policy learning and discrete SFC selection in a principled and scalable manner.

\begin{algorithm}[htbp]
\caption{SAG-MADDPG Algorithm}
\begin{algorithmic}[1]

\FOR{$\text{episode} = 1$ \TO $N_{EP}$}
    \STATE  Each destination aircraft is initialized with an initial observation $o_i(t)$.
    \FOR{$t = 1$ \TO $N_{step}$}
        \FOR{agent $i = 1$ to $M_D$}
            \STATE Obtains probability distribution vector $\breve{\mathbf{a}} _i = \pi_i(o_i^k\,|\,\theta^{\pi_i}) + \zeta$.
            \STATE Takes action $a_i$.
        \ENDFOR

        \STATE Execute actions $\mathbf{a} = \{a_1, \dots, a_{M_D}\}$, receives reward $\mathbf{r} = \{r_1, \dots, r_{M_D}\}$ by (\ref{eq22}), and the next state $\mathbf{s}(t+1) = \{o_1(t+1), \dots, o_{M_D}(t+1)\}$.        
        \STATE Store experience $\left \langle\mathbf{s}, \mathbf{a}, \mathbf{r}, \mathbf{s}(t+1)\right \rangle$ in $\mathcal{D}$
        
        \STATE Randomly select $N_{mb}$ samples from the experience pool to create a mini-batch.
            
        \FOR{agent $i = 1$ to $M_D$}
            \STATE Update actor and critic networks via (\ref{eq13}) and (\ref{eq16}), respectively.  
            \STATE Soft update target networks via (\ref{eq17}).
        \ENDFOR
        
        \STATE $\mathbf{s}(t) \leftarrow \mathbf{s}(t+1))$
    \ENDFOR
\ENDFOR
\end{algorithmic}
\end{algorithm}

\subsection{SAG-MADDPG Algorithm}

Building upon the previously defined state space, action space, and reward function, we propose a multi-agent algorithm named SAG-MADDPG, based on the MADDPG framework, as outlined in Algorithm 1. The overall workflow of Algorithm 1 is illustrated in Fig. \ref{fig:maddpg}. 

As illustrated in Fig. \ref{fig:maddpg} and detailed in Algorithm 1, at the beginning of each slot $t$, each destination aircraft obtains its initial path information $o_i(t)$ from the environment and inputs it into its actor network to generate a probability distribution vector containing all feasible paths, given by $\breve{a} _i(t) = \pi_i(o_i(t)\,|\,\theta^{\pi_i}) + \zeta$, where $\zeta$ denotes the exploration noise, and then selects the final path based on this. Once all destination aircraft have chosen their paths, the joint action $\mathbf{a}(t)$ is executed, resulting in the formation of a new aviation interconnection network topology. Upon the formation of the updated topology, the QoE—corresponding to the instantaneous reward $\mathbf{r}(t)$—is calculated for each aircraft, after which the environment transitions to a new global state $\mathbf{s}(t+1)$. The experience tuple of this interaction is stored in the replay buffer $\mathcal{D}$. A mini-batch is then randomly sampled from $\mathcal{D}$ to train the actor and critic networks, following the update rules defined in (\ref{eq13}) and (\ref{eq16}). Finally, the target networks are updated using a soft update mechanism in (\ref{eq17}), and the SAGIN system proceeds to the next time step.

The computational complexity and convergence properties of SAG-MADDPG are analyzed in detail in the Appendix.

\section{SIMULATION}

Although the SAGIN architecture features a highly dynamic three-layer network, especially in the aerial and satellite layers, the predictable trajectories of LEO satellites and the fixed flight schedules of commercial aircraft under normal conditions enable the accurate estimation of aircraft positions in real-time. In this section, we evaluate the SAG-MADDPG algorithm using real-world flight data, including global commercial flight trajectories, 20 terrestrial Internet gateway stations along the North Atlantic coast, and 66 Iridium LEO satellites. Satellite trajectories were simulated with STK software. A 24-hour period from September 17 to 23, 2017, was analyzed, selecting the 17:22 UTC slot on September 18 due to peak aircraft density \cite{refOAG}.

Flight paths were modeled as great circle arcs between airports, with aircraft positions calculated from longitude, latitude, altitude, heading, and speed, assuming a cruising altitude of 10,000 meters. The airborne network topology was built using a distributed cone-based control algorithm \cite{ref28}, ensuring each 120° cone around an aircraft contains at least one connectable neighbor. During the chosen time, 571 aircraft over the North Atlantic formed the airborne layer, classified as relay or destination nodes. The simulation used 100-200ms time slots, with each service cycle spanning three slots (300-600ms), requiring all data transmissions to complete within this period. Algorithm 1 was used to generate SFCs for destination aircraft over 4 service cycles (12 time slots).

The A2A and A2G spectrum sharing model in \cite{ref29} is adopted, allowing the A2A link to utilize idle A2G bandwidth. Due to the heterogeneous SAGIN architecture, communication parameters differ across A2A, A2G, and A2S links. Table \ref{tab:threelink} summarizes these parameters, with additional settings in Table \ref{tab:parameter}.

\begin{table}[ht]
  \caption{Distinct simulation parameters for A2G, A2A, A2S}
  \centering
  \begin{tabular}{cccc}
  \toprule
  \makecell[c]{\textbf{Parameter}} & \textbf{A2G} & \textbf{A2A} & \textbf{A2S} \\
  \midrule
  Carrier frequency $f$ (Hz) & 137M & 137M & 1.62G \\
  Transmitting antenna gain $g$dB & 3.98 & 20.97 & 21.99 \\
  Transmit power $p_t$ (dBW) & 17.8 & 12.4 & 11.8 \\
  Noise variance $\sigma^2$ (dBm/Hz) & -164.88 & -164.88 & -174.99 \\
  Maximum coverage radius $R^{max}$(nmi) & 200 & 250 & 1350 \\
  \bottomrule
  \end{tabular}
  \label{tab:threelink}
\end{table}

\begin{table}[htbp]
  \caption{The remaining simulation parameters}
  \centering
  \begin{tabular}{cc}
  \toprule
  \makecell[c]{\textbf{Parameter}} & \textbf{Value} \\
  \midrule
  Sector type & En-Route \\
  Air space & trans-Atlantic \\
  Number of IGSs on shore & 20 \\
  Number of total aircraft & 571  \\
  Number of LEO satellites & 66 \\
  Flight altitude(km) & 10  \\
  Pathloss exponent & 2 \\
  Flight velocity (km/h) & 900  \\
  The number of destination aircraft & 5-35 \\
  Buffer of the IGS (Mbit) & 300-800\\
  Buffer of the relay aircraft (Mbit) & 100-500\\

  \bottomrule
  \end{tabular}
  \label{tab:parameter}
\end{table}

In (\ref{eq8}), the coefficients $\rho_1$, $\rho_2$, and $\rho_3$ denote the weights assigned to delay, jitter, and content similarity, respectively. By adjusting these parameters, the characteristics of different types of data streams in practical aviation networks can be reflected. To reflect varying service characteristics, this study considers video, image, and text as representative data types. Similarity evaluations are conducted using publicly accessible pretrained models, with detailed parameter settings and evaluation approaches summarized in Table~\ref{tab: data types}.

\begin{table}[ht]
  \caption{Data types and characteristics}
  \centering
  \begin{tabular}{ccc}
  \toprule
  \makecell[c]{\textbf{Data type}} & \textbf{Weighting coefficients} & \textbf{Similarity model}  \\
  \midrule
  Video & $\rho_3>\rho_1,\rho_2$ (similarity sensitive)& SSIM \cite{ref25}\\
  Image & $\rho_2>\rho_1,\rho_3$ (jitter sensitive)& ResNet \cite{ref31}\\
  Text & $\rho_1>\rho_2,\rho_3$ (delay sensitive)& BERT \cite{ref24}\\
  \bottomrule
  \end{tabular}
  \label{tab: data types}
\end{table}

\subsection{Convergence of SAG-MADDPG}

\begin{figure}[htbp]
    \centering
    \includegraphics[width=0.7\linewidth]{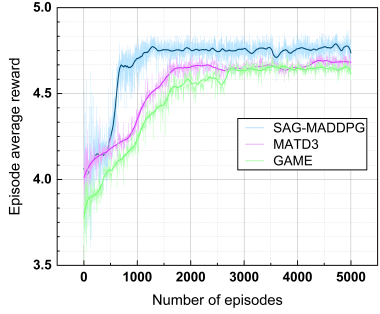}
    \caption{~~Convergence of SAG-MADDPG}
    \label{fig:convergence}
\end{figure}

\begin{figure}[htbp]
    \centering
     \includegraphics[width=0.7\linewidth, keepaspectratio]{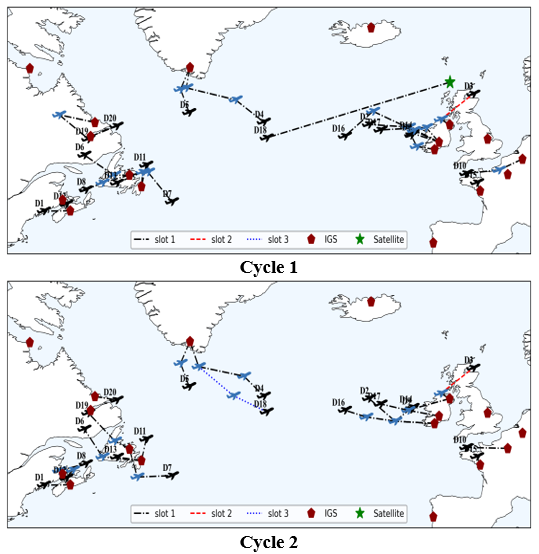}
    \caption{~~Network architecture of SAGIN}
    \label{fig:network}
\end{figure}%
    
  In this section, we evaluate the convergence performance of the proposed algorithm and compare it with the MATD3 algorithm. We also present the optimal network architecture identified in the SAGIN system. As shown in Fig. \ref{fig:convergence}, the reward curve during the training process of the SAG-MADDPG algorithm indicates that convergence is approximately achieved after 1500 episodes. The learning rates are set to $\alpha = 10^{-4}$ for the actor network and $\beta = 10^{-3}$ for the critic network. During the initial 200 episodes, the reward distribution is highly scattered due to a deliberately high exploration rate, which encourages agents to explore a broader range of actions. Although the reward curve exhibits fluctuations even in the convergence phase, these are primarily attributed to stochastic factors in the transmission process, such as interference, noise, and Doppler shifts. As a result, even with identical SFCs, performance metrics may vary across episodes, leading to fluctuations in reward values. Nevertheless, despite these fluctuations, the proposed algorithm exhibits superior learning efficiency and stability. In particular, it achieves faster convergence and a higher QoE compared with MATD3 and game-theoretic \cite{refMATD3}. The delayed policy update mechanism employed by MATD3 leads to asynchronous learning among multiple aircraft, thereby reducing the efficiency of collaborative learning \cite{refsync-update}.
  
  Fig. \ref{fig:network} illustrates the optimal network architecture obtained by the SAG-MADDPG algorithm across two consecutive cycles for a scenario with 20 destination aircraft. It can be observed that the optimal topology varies from cycle to cycle, reflecting the dynamic nature of aircraft and satellite positions over time.

\subsection{Impact of Service Type}

\begin{figure*}[ht]
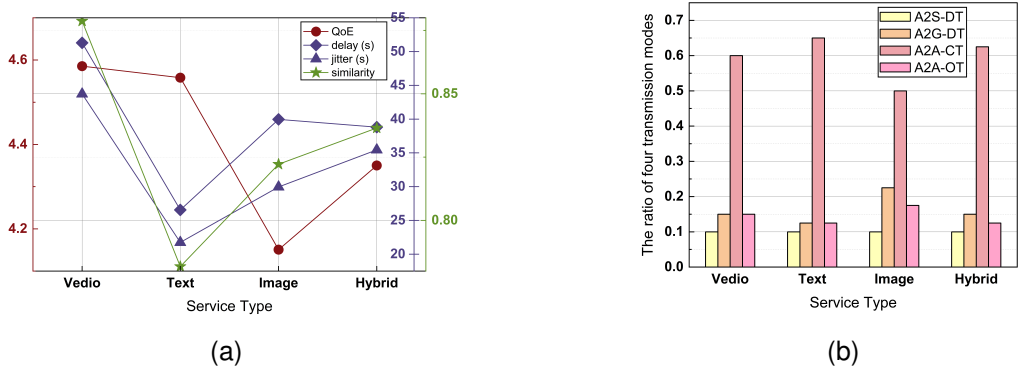

    \centering
    \subfloat[]{\includegraphics[width=2.5in]{figure/service_type/indicators.png}%
    \label{fig:service a}}
    \hfil
    \subfloat[]{\includegraphics[width=2.5in]{figure/service_type/4mode.png}%
    \label{fig:service b}}
    \caption{~~QoE metrics comparison of SAG-MADDPG under varying service types. $M_D=20$. The buffer size of IGS is 500 Mb, and the buffer size of the relay aircraft is 300 Mb. (a) QoE, delay, jitter and similarity versus network service type. (b) Ratio of A2S-DT, A2G-DT, A2A-CT, A2A-OT transmission modes.}
    \label{fig:service type}
\end{figure*}

We first analyze the impact of different service types on the overall QoE of the SAGIN network. Specifically, we compare scenarios involving the exclusive transmission of video, images, and text across the network, as well as a scenario with mixed transmission of all three data types.

Fig. \ref{fig:service type}(a) shows that video services, despite incurring higher latency and jitter due to large data sizes, still achieve high QoE thanks to strong content similarity. In contrast, text services, characterized by small packet sizes, experience lower latency and jitter, resulting in high QoE. Image services exhibit the lowest QoE overall, likely due to frequent link switching in the highly dynamic network, which introduces considerable jitter. Mixed data transmission maintains moderate values across all metrics, balancing the diverse requirements of different service types. This suggests that user-perceived quality depends on a holistic assessment of multiple performance metrics rather than a single factor. Fig. \ref{fig:service type}(b) further shows that the A2A-CT ratio is highest for text data and lowest for image data, indicating that A2A-CT is favorable for low-latency transmission but less stable due to the need for multi-VNF coordination.

\subsection{Comparison with AGIN}

\begin{figure*}[ht]
    \centering
    \subfloat[]{\includegraphics[width=1.7in]{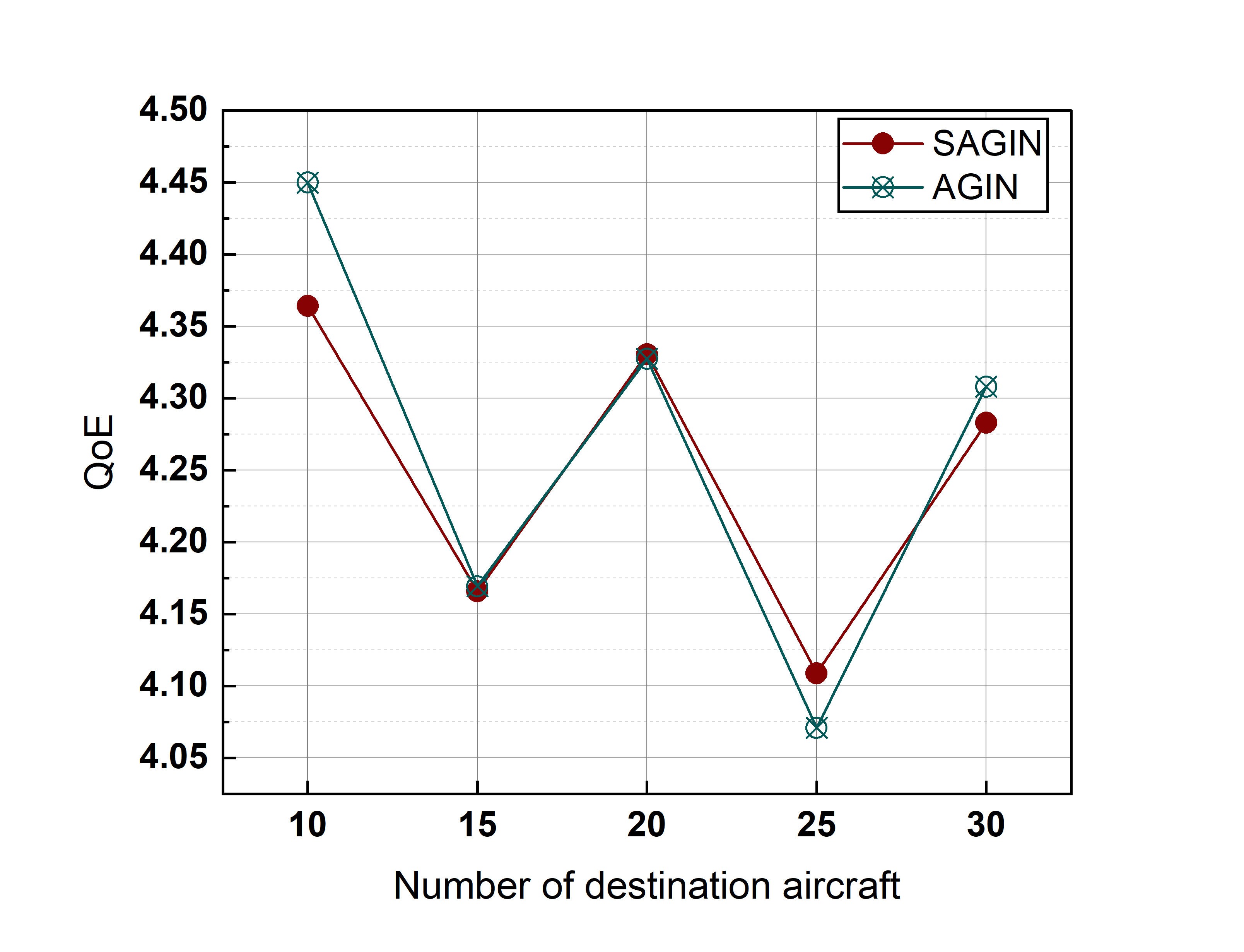}%
    \label{fig:AGINa}}
    \hfil
    \subfloat[]{\includegraphics[width=1.7in]{figure/AGIN/delay.png}%
    \label{fig:AGINb}}
    \hfil
    \subfloat[]{\includegraphics[width=1.7in]{figure/AGIN/jitter.png}%
    \label{fig:AGINc}}
    \hfil
    \subfloat[]{\includegraphics[width=1.7in]{figure/AGIN/similarity.png}%
    \label{fig:AGINd}}

    \vskip\baselineskip
    \subfloat[]{\includegraphics[width=2.1in]{figure/AGIN/5path.png}%
    \label{fig:AGINe}}
    \hfil
    \subfloat[]{\includegraphics[width=2.1in]{figure/AGIN/3mode.png}%
    \label{fig:AGINf}}
    \hfil
    \subfloat[]{\includegraphics[width=2.1in]{figure/AGIN/connect_rate.png}%
    \label{fig:connect_rate}}
    \caption{~~Comparison between AGIN and SAGIN network. The network service type is hybrid data transmission. The buffer size of IGS is 500 Mb, and the buffer size of the relay aircraft is 300 Mb. (a) QoE for SAGIN and AGIN versus the number of destination nodes. (b) Delay for SAGIN and AGIN versus the number of destination nodes. (c) Jitter for SAGIN and AGIN versus the number of destination nodes. (d) Similarity for SAGIN and AGIN versus the number of destination nodes. (e) Ratio of 1-hop, 2-hop, 3-hop, 4-hop, 5-hop chains in AGIN. (f) Ratio of A2G-DT, A2A-CT, and A2A-OT transmission modes in AGIN. (g) Coverage capabilities of A2A and A2G links.} 
    \label{fig:AGIN}
\end{figure*}

We first evaluate the connectivity of the Air–Ground–Internet (AGIN) network considering only A2G and A2A links, as illustrated in Fig. \ref{fig:AGIN}(g). Simulations are conducted using flight data from 17:22 to 17:31 UTC, with a maximum of five hops per chain. In each round, 50 destination aircraft are randomly selected, and the minimum hop count to reach an IGS is recorded over 20 repetitions. As shown in Fig. \ref{fig:AGIN}(g), approximately 88\% of aircraft can reach an IGS within three hops, while about 2\% fail to establish valid connections due to isolation and the lack of nearby relay nodes. Trajectory analysis indicates that such isolation may persist over time, leading to prolonged connectivity disruptions.

We then compare the performance of SAG-MADDPG under AGIN and SAGIN scenarios. In AGIN, three transmission modes are supported—A2G-DT, A2A-CT, and A2A-OT—with each chain limited to five hops. Network planning is performed over 12 consecutive time slots. Due to inherent topological constraints, some aircraft remain unreachable within five hops; therefore, only aircraft capable of reaching a ground station within this limit are included. To further reduce complexity, the search for 4-hop and 5-hop chains is restricted to aircraft unreachable within three hops.

Fig. \ref{fig:AGIN} compares QoE, delay, jitter, and similarity under AGIN and SAGIN for different destination-node counts. As shown in Fig. \ref{fig:AGIN}(a), the average QoE of both architectures is comparable, with differences below 0.05 except when $M_D = 10$. As illustrated in Fig. \ref{fig:AGIN}(b–d), AGIN achieves lower delay due to the absence of high-latency satellite links, whereas SAGIN exhibits lower jitter, benefiting from the greater stability of A2S transmissions. In contrast, AGIN suffers higher jitter caused by unstable multi-hop A2A relaying, where frequent relay changes amplify delay variations.

The QoE advantage of SAGIN at $M_D = 10$ is further explained in Fig. \ref{fig:AGIN}(e): approximately 25\% of the chains involve four or five hops, which in SAGIN are predominantly A2S-based, providing more stable albeit slower transmissions. Fig. \ref{fig:AGIN}(f) additionally highlights the effective utilization of A2A-CT links to support high-quality connectivity demands.

\subsection{Impact of the Number of Destination Aircraft}

\begin{figure*}[ht]
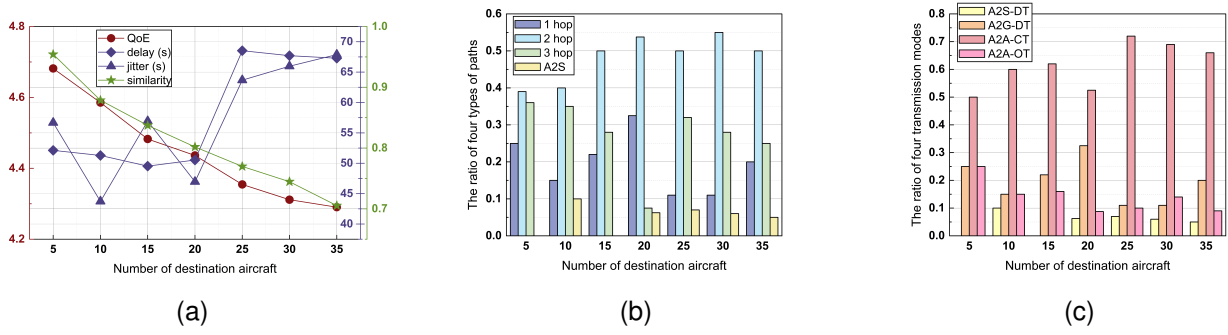

    \centering
    \subfloat[]{\includegraphics[width=2.1in]{figure/des_num/indicators.png}%
    \label{fig:desa}}
    \hfil
    \subfloat[]{\includegraphics[width=2.1in]{figure/des_num/4path.png}%
    \label{fig:desb}}
    \hfil
    \subfloat[]{\includegraphics[width=2.1in]{figure/des_num/4mode.png}%
    \label{fig:desc}}
    \caption{~~QoE metrics of SAG-MADDPG under different values of $M_D$. The network service type is pure video data transmission. The buffer size of IGS is 500 Mb, and the buffer size of the relay aircraft is 300 Mb. (a) QoE, delay, jitter, and similarity versus the number of destination aircraft. (b) Ratio of 1-hop, 2-hop, 3-hop, and A2S chain. (c) Ratio of A2S-DT, A2G-DT, A2A-CT, A2A-OT transmission modes.}
    \label{fig:destination}
\end{figure*}

We examine how various performance metrics in the SAG-MADDPG algorithm vary with the number of receiver aircraft. To ensure generality, $M_D$ receiver aircraft were randomly selected within 60°W–2°E longitude and 25°S–64°N latitude over the North Atlantic.

As shown in Fig. \ref{fig:destination}(a), increasing the number of receiver aircraft leads to lower similarity, reducing QoE in the similarity-sensitive SAGIN. This decline is due to degraded channel conditions and longer transmission distances, as more destinations reduce the number of available relay aircraft $M_R$. Shared relays and longer hops increase interference and latency. Fig. \ref{fig:destination}(b) shows that 2-hop and 3-hop SFCs dominate, highlighting the role of mutual relaying in extending communication range and maintaining quality. Fig. \ref{fig:destination}(c) further confirms the central role of A2A links, which constitute a significant portion of total transmissions.

\subsection{Impact of IGS Buffer Size}

\begin{figure*}[ht]
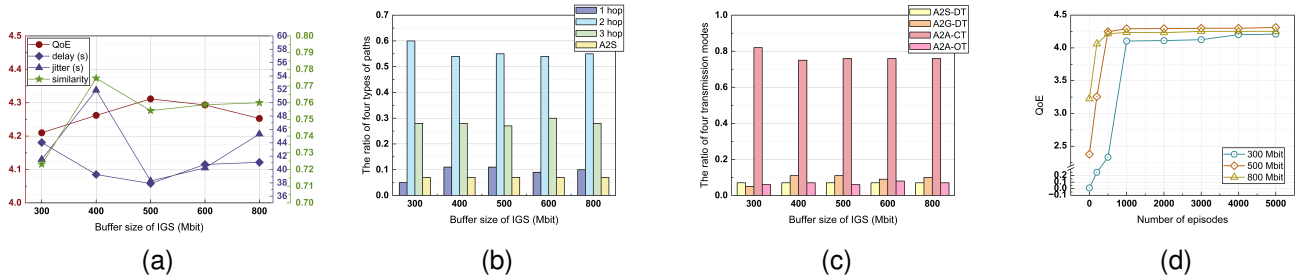

    \centering
    \subfloat[]{\includegraphics[width=1.7in]{figure/IGS_buffer/indicators.png}%
    \label{fig:igsqoe}}
    \hfil
    \subfloat[]{\includegraphics[width=1.7in]{figure/IGS_buffer/4path.png}%
    \label{fig:igspath}}
    \hfil
    \subfloat[]{\includegraphics[width=1.7in]{figure/IGS_buffer/4mode.png}%
    \label{fig:igsmode}}
    \hfil
    \subfloat[]{\includegraphics[width=1.7in]{figure/IGS_buffer/reward_curve.png}%
    \label{fig:igscurve}}
    \caption{~~QoE metrics of SAG-MADDPG under varying IGS buffer sizes. $M_D=25$. The buffer size of the relay aircraft is 300 Mb. The network service type is pure video data transmission. (a) QoE, delay, jitter, and similarity versus the size of the IGS buffer. (b) Ratio of 1-hop, 2-hop, 3-hop, and A2S chain. (c) Ratio of A2S-DT, A2G-DT, A2A-CT, A2A-OT transmission modes. (d) Iterative numbers of achieving the Nash equilibrium.}
    \label{fig:igs}
\end{figure*}

We evaluate how varying the base station buffer size $\text{BS}^{max}_{I}$ affects SAG-MADDPG performance and convergence in a similarity-sensitive network with high data demand.

As shown in Fig. \ref{fig:igs}(a), average QoE first increases and then decreases with buffer size. A small buffer limits service capacity at popular base stations, forcing aircraft to reroute frequently and lowering QoE. In contrast, an excessively large buffer increases simultaneous connections, leading to higher interference and degraded performance. Fig. \ref{fig:igs}(b) and \ref{fig:igs}(c) show that with a 300Mb buffer, fewer A2G-DT links and more 2-hop chains occur, indicating that buffer constraints push aircraft toward indirect transmissions. Fig. \ref{fig:igs}(d) shows convergence trends: most configurations converge in 1000 rounds. Notably, $\text{BS}^{max}_{I} = 500$Mb achieves faster, more stable convergence, while 300Mb requires more exploration.

\subsection{Impact of Relay Buffer Size}

\begin{figure*}[ht]
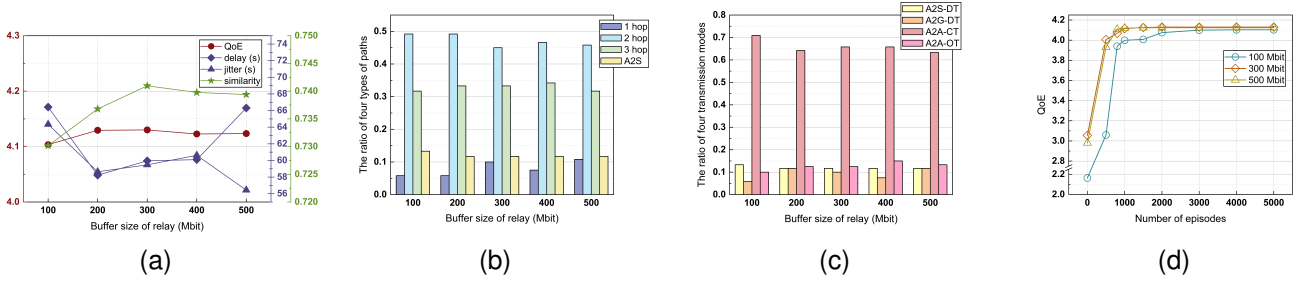

    \centering
    \subfloat[]{\includegraphics[width=1.7in]{figure/air_buffer/indicators.png}%
    \label{fig:aira}}
    \hfil
    \subfloat[]{\includegraphics[width=1.7in]{figure/air_buffer/4path.png}%
    \label{fig:airb}}
    \hfil
    \subfloat[]{\includegraphics[width=1.7in]{figure/air_buffer/4mode.png}%
    \label{fig:airc}}
    \hfil
    \subfloat[]{\includegraphics[width=1.7in]{figure/air_buffer/reward_curve.png}%
    \label{fig:aird}}
    \caption{~~QoE metrics of SAG-MADDPG under varying relay buffer sizes. $M_D=35$. The network service type is pure video data transmission. The buffer size of IGS is 500 Mb, and the buffer size of the relay aircraft is 300 Mb. (a) QoE, delay, jitter, and similarity versus the relay buffer size. (b) Ratio of 1-hop, 2-hop, 3-hop, and A2S chain. (c) Ratio of A2S-DT, A2G-DT, A2A-CT, A2A-OT transmission modes. (d) Iterative numbers of achieving the Nash equilibrium.}
    \label{fig:air}
\end{figure*}

This section examines how relay aircraft buffer size ($\text{BS}^{max}_{R}$) affects overall performance and the convergence of SAG-MADDPG. To highlight its influence, more destination nodes are selected from aircraft farther offshore with fewer nearby nodes.

As shown in Fig. \ref{fig:air}(a), QoE varies only slightly with buffer size, as destination aircraft ($M_D = 35$) account for less than 10\% of the total, and most are in dense regions with multiple chains. However, Fig. \ref{fig:air}(b) and (c) show that with $\text{BS}^{max}_{R} = 100$Mb, the use of A2S links increases, indicating insufficient relay buffer capacity limits A2A-OT transmissions to ground stations. Fig. \ref{fig:air}(d) further reveals that smaller buffers lead to longer convergence times, while oversized buffers offer no significant performance benefit.

\subsection{Results Summary}

By summarizing and analyzing the experimental results, the following conclusions can be drawn:

\begin{enumerate}
\item{} When the total number of nodes in the SAGIN network is approximately 650, and the number of destination nodes ranges from 5 to 35, all variants of the proposed Algorithm SAG-MADDPG can converge to a Nash equilibrium within 1500 iterations. SAG-MADDPG with the same network settings achieves better QoE than MATD3 and the game.
\item{} Integrating ground, satellite, and airborne networks is essential for reliable communication services in civil aviation. A2A communications take advantage of the mobility of airborne networks to provide flexible connectivity among aircraft, while A2S communications ensure wide-area access and coverage. 
\item{} Variations in service types lead to different user sensitivities toward key performance indicators, resulting in diverse QoE outcomes. To ensure a satisfactory user experience, it is essential to allocate network resources flexibly and dynamically in accordance with the specific requirements of each service.
\item{} The buffer capacity of VNFs has a significant impact on network formation and performance, and both insufficient and excessive buffering can be detrimental. For 5–35 destination nodes, a base station buffer of 500 Mb is optimal. Similarly, while too small a too-small relay buffer limits opportunistic transmission (OT), increasing it beyond 300 Mb yields no further performance gains under the current settings.
\end{enumerate}

\section{CONCLUSION}

This paper proposes a three-layer heterogeneous SAGIN architecture to enable seamless in-flight Internet services in remote areas without terrestrial infrastructure. By integrating multiple transmission links and Service Function Chains under the NFV framework, the proposed system effectively supports heterogeneous service demands with limited communication resources.

A tailored SAG-MADDPG algorithm is developed to solve the resulting multi-task SFC scheduling problem with the objective of maximizing overall Quality of Experience. Simulation results based on real flight data show that the proposed algorithm converges reliably within a finite number of iterations and consistently outperforms benchmark schemes in terms of QoE. These results demonstrate the effectiveness and robustness of the proposed framework for large-scale and highly dynamic aviation networking scenarios.

With the continued growth of the aviation tourism industry and a surge in airborne users, more computational tasks will emerge at aircraft terminals. Future research may explore task offloading and edge processing strategies at the terminal level to further enhance service quality and network efficiency.

\appendix
\section*{Computational Complexity and Convergence Analysis of SAG-MADDPG}
\label{app:complexity_convergence}

In this appendix, we analyze the computational complexity and convergence
properties of the proposed SAG-MADDPG algorithm.

\subsection{Computational Complexity}

Consider a SAGIN with $M_D$ destination aircraft, each modeled as an agent equipped with an actor–critic pair under the MADDPG framework. Let $\mathbf{P}_{\max}^{\mathrm{dim}}$ denote the maximum number of feasible SFCs retained after LCSS-based filtering, and let $N_{\mathrm{mb}}$ be the mini-batch size.

In each training iteration, the dominant computational overhead arises from forward and backward propagation in the actor and critic networks. The overall per-iteration complexity, therefore, scales as $\mathcal{O}\!\left(M_D \cdot N_{\mathrm{mb}} \cdot \mathbf{P}_{\max}^{\mathrm{dim}}\right)$.

The LCSS-based SFC filtering incurs an additional cost of $\mathcal{O}(|\mathbf{P}_i|^2 \cdot L)$ per agent, where $|\mathbf{P}_i|$ is the original candidate set size and $L$ is the average SFC length. Since this procedure is triggered only upon significant topology changes in the reconfigurable time-expanded graph, its amortized cost is negligible compared with the online training complexity.

Overall, SAG-MADDPG exhibits polynomial-time complexity and scales linearly with $M_D$, making it suitable for large-scale SAGIN deployments.

\subsection{Convergence Properties}

The SAG-MADDPG algorithm follows the deterministic policy gradient framework under a centralized training and decentralized execution (CTDE) paradigm. During training, each agent employs a centralized critic with access to global state and joint actions, effectively mitigating the non-stationarity induced by concurrent multi-agent learning.

Under standard assumptions in deep reinforcement learning—such as bounded rewards, sufficiently small learning rates, and expressive function approximators—the deterministic policy gradient method converges to a locally optimal policy. In the cooperative multi-agent setting considered here, the centralized critic guides learning toward a local Nash equilibrium of the SAGIN topology optimization game.

Moreover, experience replay and soft target-network updates reduce gradient variance and smooth parameter evolution, significantly enhancing training stability. Simulation results in Section V confirm this analysis: training rewards converge within a finite number of episodes, and SAG-MADDPG achieves faster convergence and higher QoE than the MATD3 baseline.

\bibliographystyle{IEEEtran}
\bibliography{refs}

\end{document}